\documentclass[useAMS,usenatbib,fleqn]{mn2e}

\usepackage{graphicx}
\usepackage{textcomp}
\usepackage{amsmath}
\usepackage{breqn}
\usepackage{wasysym}
\usepackage{amssymb}

\title[Refraction in planetary atmospheres]{Refraction in planetary atmospheres: improved analytical expressions and comparison with
a new ray tracing algorithm}
\author[Yan B\'etr\'emieux and Lisa Kaltenegger]{Yan B\'etr\'emieux$^{1}$ \thanks{e-mail: betremieux@mpia.de} and 
Lisa Kaltenegger$^{1,2}$\thanks{Now at the Institute for Pale Blue Dots, Cornell University} \\
$^{1}$Max-Planck-Institut f\"{u}r Astronomie, K\"{o}nigstuhl 17, 69117 Heidelberg, Germany\\
$^{2}$Harvard-Smithsonian Center for Astrophysics, 60 Garden street, Cambridge MA 02138, USA}
\begin{document}


\pagerange{\pageref{firstpage}--\pageref{lastpage}} \pubyear{2015}

\maketitle

\label{firstpage}

\begin{abstract}

Atmospheric refraction affects to various degrees exoplanet transit, lunar eclipse, as well as stellar occultation observations.
Exoplanet retrieval algorithms often use analytical expressions for the column abundance along a ray traversing the 
atmosphere as well as for the deflection of that ray, which are first order approximations valid for low densities 
in a spherically symmetric homogeneous isothermal atmosphere. We derive new analytical formulae for both 
of these quantities, which are valid for higher densities, and use them to refine and validate a new ray tracing 
algorithm which can be used for arbitrary atmospheric temperature-pressure profiles. We illustrate with simple
isothermal atmospheric profiles the consequences of our model for different planets: temperate Earth-like and Jovian-like 
planets, as well as HD189733b, and GJ1214b. We find that, for both hot exoplanets, our treatment of refraction does 
not make much of a difference to pressures as high as 10~atmosphere, but that it is important to consider the variation 
of gravity with altitude for GJ1214b. However, we find that the temperate atmospheres have an apparent scale height 
significantly smaller than their actual density scale height at densities larger than 1~amagat, thus increasing the difficulty 
of detecting spectral features originating in these regions. These denser atmospheric regions form a refractive boundary 
layer where column abundances and ray deflection increases dramatically with decreasing impact parameter. This refractive 
boundary layer mimics a surface, and none of the techniques mentioned above can probe atmospheric regions denser 
than about 4~amagat on these temperate planets.

\end{abstract}


\section{Introduction}\label{intro}

Most analytical expressions of an exoplanetary primary transit lightcurve
are derived from analytical expressions of the optical depth along
an unrefracted ray, i.e. that travels in a straight light, 
through a homogeneous isothermal atmospheres
(see e. g. Fortney 2005; Lecavelier des Etangs et al. 2008; Benneke \& Seager 2012; 
Howe \& Burrows 2012; de Wit \& Seager 2013; Griffith 2014). However, several papers 
(Hui \& Seager 2002; Sidis \& Sari 2010; Garc\'ia Mu\~noz et al. 2012; 
B\'etr\'emieux \& Kaltenegger 2013, 2014; Misra, Meadows \& Crisp 2014; 
Misra \& Meadows 2014) have emphasised the importance of the refractive bending 
of light by exoplanetary atmospheres on these lightcurves. Refraction is 
also important in various observational geometries in our Solar System. It influences 
the perceived brightness of a star during a stellar occultation by a planetary atmosphere
(see the review by Smith \& Hunten~1990) which can then be used to infer the planet's 
atmospheric composition. It also determines the brightness of moons eclipsed by their 
parent planet. Indeed, as a moon moves deeper in the penumbra, solar radiation
must traverse deeper atmospheric regions of the eclipsing planet in order 
to be bent sufficiently to reach the moon. Several Lunar eclipse observations
were analysed to derive the transmission of Earth's atmosphere 
(Pall\'e et al. 2009; Vidal-Madjar et al. 2010; Garc\'ia Mu\~noz et al. 2012, Arnold 
et al. 2014) using this principle.

Given the importance of atmospheric refraction in these various observational geometries, 
it is worthwhile to have analytical formulae which include refraction, both to compute
the ray's deflection, as well as the integrated number density along the deflected 
ray, or its column abundance. Combined with an atmosphere-averaged extinction 
cross section, one can compute the optical depth along the ray, and the corresponding atmospheric
transmission with Beer's law. However, computing the column abundance and the deflection requires solving 
complicated integrals for which analytical expressions exist only under simplifying assumptions. 
In section~\ref{analytical}, we derive analytical solutions of these integrals 
by expanding the integrand in a Taylor-series up to the second order with respect to 
refractivity, and then evaluate the integral for each of these terms. This is  
done for a homogeneous isothermal atmosphere, since this is the level of complexity considered by 
some exoplanet atmosphere retrieval algorithm (see e.g. Charbonneau et al. 2009; 
Howe \& Burrows 2012; Anglada-Escud\'e et al. 2013; Ehrenreich et al. 2014; Waldmann et al. 2014).
We also derive an order-of-magnitude estimate of the sum of the remaining uncomputed higher-order 
terms in the Taylor series, in order to determine the density region over which our analytical 
expressions are useful.

Since planetary atmospheres are usually not isothermal, we have to rely on numerical 
methods to deal with these general cases for which analytical solutions do not exist. 
In section~\ref{numerical}, we describe MAKEXOSHELL, a new numerical ray tracing algorithm that computes 
the column abundance and the deflection for rays traversing a spherically symmetric atmosphere given 
an arbitrary one-dimensional (1-D) temperature-pressure profile as input. We also describe how we 
compute the model atmosphere, as well as its refractivity, which are inputs to MAKEXOSHELL. 
In section~\ref{validation}, we compare the output of the numerical algorithm 
with our analytical expressions for an Earth-like planet with a homogeneous isothermal atmosphere, in
order to mutually validate both methods. We also discuss the impact of our new model 
on temperate Earth-like and Jovian-like exoplanets, as well as for two already well-observed 
exoplanets: HD189733b and GJ1214b. We summarise our results in section~\ref{conclusion}.

\begin{figure}
\includegraphics[scale=0.30]{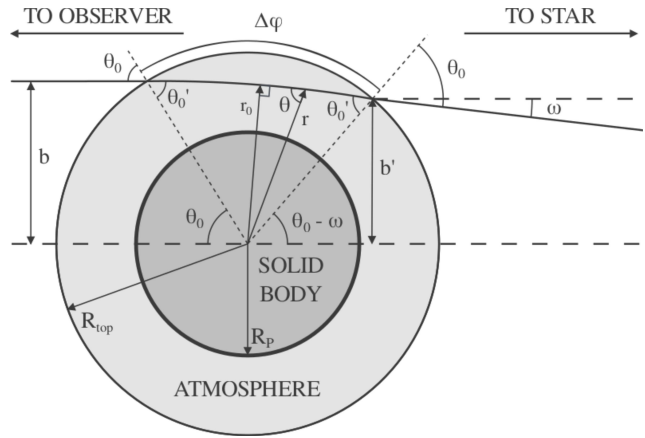}
\caption{Trajectory through a planetary atmosphere of an observed light ray during 
an exoplanetary primary transit. The solid body, or opaque region, of the planet (dark grey) and 
its atmosphere (light grey) are shown, along with the axis running from the observer to the centre of the 
planet (dashed line) with the observer to the left and the star to the right. The radius of the planetary 
surface, $R_p$, and of the top of the atmosphere, $R_{top}$, as well as the zenith angle, $\theta$, of 
the ray for a given height, r, are also indicated. A light ray observed with an impact parameter~$b$, 
entered the atmosphere with an incidence angle $\theta_0$ a projected distance $b'$ to the centre 
of the planet with respect to the observer, reached a grazing height $r_0$, and is deflected 
by $\omega$ by the atmosphere.
\label{fig1}}
\end{figure}

\section{Analytical expressions}\label{analytical}

\subsection{Basic equations}

As light rays traverse a planetary atmosphere, they are bent by refraction 
toward the surface of the planet (see Fig.~\ref{fig1}) due to the exponential 
increase of atmospheric refractivity (the refraction index minus one), 
with decreasing altitude. The trajectory of a ray is described by an 
invariant, $b$, which is equal to its impact parameter 
(Phinney \& Anderson 1968; B\'etr\'emieux \& Kaltenegger 2013, 2014)
\begin{equation}\label{raypath}
b = r (1 + \nu) \sin{\theta} = r_0 (1 + \nu_0) = R_{top} \sin{\theta_0} .
\end{equation}
Here, $r$ is the radial position with respect to the centre of the planet (i.e. radius + altitude), 
which for the sake of brevity we will refer to as height throughout the paper, 
while $\nu$ and $\theta$ are the refractivity of the atmosphere and the zenith angle of the
ray at that height, respectively. $r_0$ is the minimum height reached by a ray, or its grazing height,
where the atmospheric refractivity is $\nu_0$. $R_{top}$ is the radial position of the top of the 
atmosphere and $\theta_{0}$ is the zenith angle of the ray at the top of the atmosphere, or simply 
its incidence angle. The atmosphere deflected the ray by $\omega$, and the ray entered the atmosphere
a projected distance $b'$ to the centre of the planet, with respect to the observer, given by
\begin{equation}
b' = \left( \frac{\sin(\theta_0 - \omega)}{\sin{\theta_0}} \right) b .
\end{equation}

The change in zenith angle of the ray with height is given by, 
\begin{equation}\label{gradzenith}
 \frac{d\theta}{dr} = - \frac{\tan{\theta}}{(1 + \nu)} \left( \frac{d\nu}{dr} \right) - \frac{\tan\theta}{r} .
\end{equation}
The first term is the refractive term and comes from Snell's law. The second term is the geometric term, and 
comes from the equation of an unrefracted ray, $b = r \sin\theta$, which proceeds in a straight line.
Note that all of the equations involving angles are
valid when these are expressed in radians, even though in section~\ref{validation} we quote results in degrees.

The atmospheric refractivity is given by,
\begin{equation}\label{refrac}
\nu = \left( \frac{n}{n_{STP}} \right) \sum_{j} f_{j} {\nu_{STP}}_{j} = \left( \frac{n}{n_{STP}} \right) \nu_{STP} ,
\end{equation}
where $n$ is the number density, $n_{STP}$ is the number density at standard temperature and pressure
(STP) also known as Loschmidt's number, $f_{j}$ is the mole fraction of 
the j$^{th}$ chemical species, ${\nu_{STP}}_{j}$ is the STP refractivity of the j$^{th}$ species, while $\nu_{STP}$ 
is the STP refractivity of the atmosphere. The ratio inside the parenthesis in equation~\ref{refrac}
is the number density expressed in units of amagat. Note that the STP refractivity is simply
the refractivity for a density of 1~amagat, which can be achieved for 
temperatures other than 273.15~K with the appropriate pressure. 

The number density, $n$, in an atmospheric region varies with height, and this dependence can 
be expressed by
\begin{equation}\label{expoatmos}
n_U = n_L e^{-(r_U - r_L)/H } ,
\end{equation}
so that
\begin{equation}
\frac{dn}{dr} = -\frac{n}{H} ,
\end{equation}
where $H$ is the density scale height in that atmospheric region, 
and the $U$ and $L$ subscripts denote quantities for the upper and lower boundaries of the region, respectively.
Thus, 
\begin{equation}\label{gradindex}
 \frac{d\nu}{dr} = \frac{\nu_{STP}}{n_{STP}} \left( \frac{dn}{dr} \right) 
= - \left( \frac{n}{n_{STP}} \right) \frac{\nu_{STP}}{H} = - \frac{\nu}{H} .
\end{equation}
Note that this result holds only for a homogeneous atmospheric region where the composition of the major 
constituents that contribute significantly to the refraction index does not change with height, i.e. that
$\nu_{STP}$ is constant with height. This assumption will be carried throughout all derivations.

\subsection{Lower atmospheric boundary}\label{lowerbound}

In order for a ray to traverse the atmosphere, $\theta$ must increase 
as $r$ decreases (or $d\theta$/$dr < 0$) so that a downward-going ray reaches 
a minimum height, $r_0$, where it grazes the atmosphere ($\theta = \pi/2$).
If $d\theta$/$dr > 0$, the ray will spiral deeper into the atmosphere until it is 
absorbed. Using equations~\ref{gradzenith}, we can derive that 
\begin{equation}
- \frac{r}{(1+\nu)} \left( \frac{d\nu}{dr}\right) < 1
\end{equation}
is the necessary condition for a ray to traverse the atmosphere. Using equation~\ref{gradindex}, 
this can be rewritten as,
\begin{equation}\label{condition}
\left( \frac{\nu}{1+\nu} \right) \frac{r}{H} < 1  . 
\end{equation}
In the special case when this expression is equal to one, a ray which is tangential 
to the atmosphere will circle the planet as its radius of curvature matches its height.
Knowing the height-dependence of the density, and thus refractivity, this equation can be solved 
recursively for this special height. As a ray approaches this height, both the column abundance 
and the deflection tend toward infinity. Atmospheric regions below this height cannot be probed by 
transmitted stellar radiation under any circumstance during a stellar occultation or an exoplanet 
primary transit. This is true only for a perfectly spherical planet free of horizontal density gradients (i.e. no 
weather or turbulence). Departure from these assumptions creates a diffuse boundary about which
a grazing ray can or cannot escape the planet depending on local conditions. Furthermore, 
since refractivities vary with wavelength, so does the height of this lower boundary.

This lower boundary is always located below the critical altitude, described in \citet{YB_LK_2014}, 
below which atmospheric regions cannot be probed by transmission spectroscopy when an exoplanet occults 
the central area of its host star. This critical altitude occurs where a ray, that comes from the 
opposite limb of the star, reaches an atmospheric critical density which causes it to be deflected 
by a critical deflection, $\omega_c$, toward the observer. During the course of the 
transit, the symmetry is broken and part of the planetary limb can be probed to lower altitudes while 
the opposite limb's observable region is restricted to higher altitudes. However, the lowest altitude
that one can probe only asymptotically approaches the lower boundary determined by 
equation~\ref{condition}, so that this boundary is the deepest atmospheric region that can ever be 
probed by exoplanet transmission spectroscopy or by stellar occultation.

\subsection{Column abundance and ray deflection integrals}

The column abundance, $N$, along a refracted ray, as well as its 
deflection, $\omega$, can be computed from
\begin{equation}
 N = 2 \int_{r_0}^{\infty}dN \equiv 2 \int_{r_0}^{\infty} F_1 dr \label{colab}
\end{equation}
and
\begin{equation}
\omega = 2 \int_{r_0}^{\infty}d\omega \equiv 2 \int_{r_0}^{\infty} F_2 dr \label{defl} 
\end{equation}
where the factor of 2 comes from a ray traversing an altitude region twice. Note that 
we have labelled the integrands in terms of the integration parameter $r$ in both equations 
for future use (see section~\ref{taylorexp}).

The incremental column abundance, $dN$, of a ray is given by,
\begin{equation}\label{incpath}
dN = \frac{n}{\cos{\theta}}dr . 
\end{equation}
The incremental deflection of a ray (expressed in radians), $dw$, is simply the first 
term in equation~\ref{gradzenith} (see also Goldsmith 1963;
Auer \& Standish 2000), which we rewrite as
\begin{equation}\label{incdef}
d\omega = - \frac{\tan\theta}{(1+\nu)} \left( \frac{d\nu}{dr} \right) dr = \left( \frac{\nu}{1+ \nu} \right) \frac{\tan\theta}{H} dr
\end{equation}
for a homogeneous planetary atmosphere. 

The trigonometric functions can be rewritten entirely in terms of $r$ and $\nu$.
From equation~\ref{raypath}, we have
\begin{equation}
 \sin\theta = \frac{b}{r(1+\nu)} = \frac{r_0(1+\nu_0)}{r(1+\nu)}  , 
\end{equation}
\begin{eqnarray}
 \frac{1}{\cos\theta} = \frac{1}{\sqrt{1 - \sin^2\theta}} = \left( 1- \left( \frac{r_0(1+\nu_0)}{r(1+\nu)} \right)^2 \right)^{-1/2} ,
\end{eqnarray}
and
\begin{equation}
 \tan\theta = \frac{\sin\theta}{\cos\theta} = \left( \left( \frac{r(1+\nu)}{r_0(1+\nu_0)} \right)^2 -1 \right)^{-1/2} .
\end{equation}

Equations~\ref{colab} and \ref{defl} are then given, for a homogeneous atmosphere, by
\begin{equation}
 N = 2 \int_{r_0}^{\infty} n \left( 1- \left( \frac{r_0(1+\nu_0)}{r(1+\nu)} \right)^2 \right)^{-1/2} dr , \label{colabexpr}
\end{equation}
and 
\begin{equation}
 \omega = 2 \int_{r_0}^{\infty} \left( \frac{\nu}{1+ \nu} \right) \frac{1}{H}
\left( \left( \frac{r(1+\nu)}{r_0(1+\nu_0)} \right)^2 -1 \right)^{-1/2} dr   \label{deflexpr}
\end{equation}
where most of the height-dependence is buried inside the refractivity.

\subsection{Taylor-series expansion of integrands}\label{taylorexp}

No analytical solutions exist for the complicated integrals in equations~\ref{colabexpr} and \ref{deflexpr}, 
even in the case of a homogeneous isothermal atmosphere. 
However, refractivities are typically small, on the order of $10^{-4}$ at STP, such that one can expand the integrand 
in both equations in Taylor series in $\nu_0$, prior to carrying-out the integration. We carry out this Taylor
expansion for both integrands $F_1$ and $F_2$ to the second order such that
\begin{eqnarray}\label{tscolabun}
 dN & = & dN_0 + dN_1 + dN_2 + ... \\  
    & = & F_1|_{\nu_0 = 0} + \nu_0 \left( \frac{dF_1}{d\nu_0} \right)_{\nu_0=0} + 
   \frac{\nu_0^2}{2} \left( \frac{d^2F_1}{d\nu_0^2} \right)_{\nu_0=0} + ...  ,  \nonumber  
\end{eqnarray}
and 
\begin{eqnarray}\label{tsdeflection}
 d\omega & = & d\omega_0 + d\omega_1 + d\omega_2 + ... \\ 
    & = & F_2|_{\nu_0 = 0} + \nu_0 \left( \frac{dF_2}{d\nu_0} \right)_{\nu_0=0} 
    + \frac{\nu_0^2}{2} \left( \frac{d^2F_2}{d\nu_0^2} \right)_{\nu_0=0} + ...  \nonumber  .
\end{eqnarray}

For a homogeneous isothermal atmosphere, we have the additional relations,
\begin{equation}
n = n_0 e^{-(r - r_0)/H }
\end{equation}
and 
\begin{equation}
\nu = \nu_0 e^{-(r - r_0)/H }
\end{equation}
where $n_0$ and $\nu_0$ are the number density and the refractivity at a ray's grazing height, respectively. 
The derivation of the various Taylor expansion terms is rather tedious and lengthy, so we will here only 
quote the results: 
\begin{equation}
 F_1|_{\nu_0 = 0} = \frac{n}{\sqrt{1-(r_0/r)^2}} = \frac{n(r/r_0)}{\sqrt{(r/r_0)^2-1}}
\end{equation}
\begin{equation}
 \left( \frac{dF_1}{d\nu_0} \right)_{\nu_0=0} = \frac{n (r/r_0) ( 1 - e^{-(r-r_0)/H} )}{ [ (r/r_0)^2-1 ]^{3/2} }
\end{equation}
\begin{multline}
 \left( \frac{d^2F_1}{d\nu_0^2} \right)_{\nu_0=0}  = \frac{3n(r/r_0)^3 ( 1 - e^{-(r-r_0)/H} ) ^2}{ [ (r/r_0)^2 -1 ]^{5/2} }  \\
  - \frac{2n(r/r_0) ( 1 - e^{-(r-r_0)/H} )} { [ (r/r_0)^2 -1 ]^{3/2} }
\end{multline}
and,
\begin{equation}
 F_2|_{\nu_0 = 0} = 0
\end{equation}
\begin{equation}
 \left( \frac{dF_2}{d\nu_0} \right)_{\nu_0=0} = \frac{1}{H} \frac{e^{-(r-r_0)/H}}{\sqrt{(r/r_0)^2-1}}
\end{equation}
\begin{multline}
 \left( \frac{d^2F_2}{d\nu_0^2} \right)_{\nu_0=0} = \frac{2}{H} \frac{e^{-(r-r_0)/H}}{\sqrt{(r/r_0)^2-1}} \times \\
   \left[ \frac{(r/r_0)^2( 1 - e^{-(r-r_0)/H} ) }{(r/r_0)^2 -1}  - e^{-(r-r_0)/H} \right] .
\end{multline}

\subsection{``Modified Bessel functions'' solution}

At first glance, it seems that we have made the problem more complicated as we have transformed 
each of our initial integrals into sums of three integrals, each of which also seem impossible to solve. 
However, a suggestion of a solution appears when one considers the modified Bessel functions
of the second kind \citep{Arfken_1985}, 
\begin{equation}\label{bessel}
K_t (z) = \frac{\pi^{1/2}}{(t - 1/2)!} \left(\frac{z}{2} \right)^t \int_1^\infty e^{-zx} (x^2 -1)^{t - 1/2} dx
\end{equation}
of which the first two functions are
\begin{equation}
K_0(z) = \int_1^\infty e^{-zx} (x^2 -1)^{- 1/2} dx ,
\end{equation}
and
\begin{equation}
K_1(z) = z \int_1^\infty e^{-zx} (x^2 -1)^{1/2} dx ,
\end{equation}
which were derived from equation~\ref{bessel}, knowing that $(-1/2)! = \pi^{1/2}$.

Consider, as a first step, the zeroth term in the column abundance integral,
\begin{eqnarray}
 N_0 & = & 2 \int_{r_0}^{\infty} dN_0 = 2 \int_{r_0}^{\infty} \frac{n(r/r_0)}{\sqrt{(r/r_0)^2-1}} dr  \nonumber \\
    & = & 2 \int_{r_0}^{\infty} \frac{n_0(r/r_0)e^{-(r-r_0)/H}}{\sqrt{(r/r_0)^2-1}} dr  .
\end{eqnarray}
With the change of variable, $x = r/r_0$, it becomes
\begin{equation}
 N_0 = 2 n_0 r_0 e^{r_0/H} \int_{1}^{\infty} \frac{xe^{-(r_0/H)x}}{(x^2-1)^{1/2}} dx 
\end{equation}
which can be integrated by parts to yield
\begin{multline}
 N_0 = \left[ 2 n_0 r_0 e^{r_0/H} (x^2-1)^{1/2}e^{-(r_0/H)x} \right]_1^{\infty} \\
          + 2 n_0 r_0 e^{r_0/H} (r_0/H) \int_{1}^{\infty} e^{-(r_0/H)x}(x^2-1)^{1/2} dx .
\end{multline}
The values of the first term at 1 and $\infty$ are both zero so the first term drops out. Whereas it
is straightforward to derive for the value of 1, the $\infty$ case requires the use of l'H\^opital's rule.
The second term can be rewritten in terms of the modified Bessel functions of the second kind, 
so that we obtain
\begin{equation}\label{col0}
 N_0 =  2 n_0 r_0 e^{r_0/H} K_1(r_0/H)  = 2 n_0 r_0 K_1^*(r_0/H)
\end{equation}
where we adopt the following short-hand notation,
\begin{equation}
 K_t^*(y) = e^y K_t(y)
\end{equation}
for the remainder of this paper. This expression for the column abundance
has been previously derived in terms of these modified Bessel functions (see e.g. Griffith 2014). 
However, this is only the leading term in our Taylor series, and corresponds to the non-refractive case.

Each of the terms in the column abundance and deflection expansion can also be rewritten 
in terms of these modified Bessel functions using the same change of variable, and then 
integrating by parts a number of times. At each integration by parts, we get a term 
that drops out when evaluated at 1 and $\infty$ and an integral term that can either 
be expressed as a modified Bessel function or must be further integrated by parts. 
Again, the derivation is straightforward but rather lengthy so we only here quote 
the results:
\begin{equation}\label{col1}
 N_1 =  2 n_0 r_0 \left(\frac{r_0 \nu_0}{H}\right) \left[ 2 K_0^*\left(\frac{2r_0}{H}\right) 
- K_0^*\left(\frac{r_0}{H}\right) \right]
\end{equation}
\begin{multline}\label{col2}
 N_2 = n_0 r_0 \left(\frac{r_0 \nu_0}{H} \right)^2 \left[ 9K_1^*\left(\frac{3r_0}{H}\right) 
       - 8K_1^*\left(\frac{2r_0}{H}\right) + K_1^*\left(\frac{r_0}{H}\right) \right] \\
     + n_0 H \left(\frac{r_0 \nu_0}{H} \right)^2 \left[ -7K_0^*\left(\frac{3r_0}{H}\right) 
       + 4K_0^*\left(\frac{2r_0}{H}\right) + K_0^*\left(\frac{r_0}{H}\right)\right]
\end{multline}
and
\begin{equation}\label{defl0}
 \omega_0 = 0
\end{equation}
\begin{equation}\label{defl1}
 \omega_1 = 2 \left( \frac{r_0 \nu_0}{H} \right)  K_0^*\left(\frac{r_0}{H}\right)
\end{equation}
\begin{multline}\label{defl2}
 \omega_2 = 2 \left( \frac{r_0 \nu_0}{H} \right)^2 \left[2 K_1^*\left(\frac{2r_0}{H}\right) 
-  K_1^*\left(\frac{r_0}{H}\right)  \right] \\
 + 2 \left( \frac{r_0}{H} \right) \nu_0^2 \left[-2 K_0^*\left(\frac{2r_0}{H}\right) 
+ K_0^*\left(\frac{r_0}{H}\right)  \right] .
\end{multline}
Note that the leading term of the Taylor-series expansion for deflection, shown in equation~\ref{defl0}, is 0 
as expected since the leading term is the non-refractive case. Without refraction, light is not bent by the 
atmosphere. 

\subsection{``General power-series'' solution}

Expressions in power-series exists for these modified Bessel functions \citep{Arfken_1985}, and are given by:
\begin{equation}\label{besselfunc}
K_t^* (y) = \sqrt{\frac{\pi}{2y}} \left[ 1 + \frac{(4t^2 - 1)}{8y} + \frac{(4t^2-1)(4t^2-9)}{128 y^2} + ... \right] .
\end{equation}
In our case, $y$ is always a multiple of $r_0/H$, which is always large. Thus, the second, third and higher-order 
terms in equation~\ref{besselfunc} decrease by successive powers of multiples of $r_0/H$, and are significantly
smaller than the leading term in this series. 

Collecting terms, equations~\ref{tscolabun} and \ref{tsdeflection} can be rewritten more generally as
\begin{equation}\label{seriescol}
 N = \sqrt{\frac{2\pi r_0}{H}} (n_0 H) \left[1 + C_0 + \sum_{j = 1}^{\infty} C_j \left( \frac{r_0\nu_0}{H} \right)^j \right] 
\end{equation}
and 
\begin{equation}\label{seriesdefl}
 \omega = \sqrt{\frac{2\pi r_0}{H}} \nu_0 \left[1 + D_0 + \sum_{j = 1}^{\infty} D_j \left( \frac{r_0\nu_0}{H} \right)^j \right] ,
\end{equation}
where $C_j$ and $D_j$ are coefficients for the column abundance and deflection, respectively. 
Even though we have only explicitly collected terms up to $j=2$ for the column abundance, and $j=1$ for the deflection, we can 
determine by inspection that higher-order terms follow this trend. These coefficients are themselves power-series expansion 
in term of $H/r_0$, but each of them quickly converges since this ratio is always small.

Both series depends on $\sqrt{2\pi r_0/H}$, the ratio of gas column abundance along a grazing ray 
to that along a vertical ray (given by $n_0 H$) for a curved exponential atmosphere. This ratio, which we will 
call the slant factor, has been described before (see e.g. Fortney 2005). The leading non-zero term in both series 
matches previous expressions of the column abundance without refraction (see references listed in section~\ref{intro}), 
as well as first-order expressions for the ray deflection (Goldsmith 1963), respectively. Higher-order terms in 
the series are surprisingly dependent on powers of $(r_0 \nu_0/H)$, and not simply $\nu_0$. Whereas $\nu_0$ is usually 
very small, it increases exponentially with depths so that combined with $r_0/H$, which is large, $(r_0 \nu_0/H)$ can become 
of order unity. Hence, higher-order terms in the series are far from negligible when a ray travels deep enough in 
a planetary atmosphere. So over what range of densities can we use our analytical expressions? 

We have only computed the coefficients up to the second order in refractivity as the length of the
calculations, and correspondingly the odd of making a mathematical mistake, increases steeply with order. 
However, we can make an order-of-magnitude estimate of the sum of the terms for which we have not explicitly 
computed the $C_j$ and $D_j$ coefficients, for both the column abundance, $N_{rest}$, and the deflection, $\omega_{rest}$.
This is done by assuming that the unknown coefficients are all identical to the highest-order computed coefficient. 
Under this crude assumption, we can write
\begin{equation}\label{gencolabun}
 N_{rest} = \sqrt{\frac{2\pi r_0}{H}} (n_0 H) C_2 \left( \frac{r_0\nu_0}{H} \right)^3 \sum_{l = 0}^{\infty}
\left( \frac{r_0\nu_0}{H} \right)^l
\end{equation}
and
\begin{equation}\label{gendeflect}
 \omega_{rest} = \sqrt{\frac{2\pi r_0}{H}} \nu_0 D_1 \left( \frac{r_0\nu_0}{H} \right)^2 \sum_{l = 0}^{\infty}
\left( \frac{r_0\nu_0}{H} \right)^l 
\end{equation}
where the leading term for $C_2$ and $D_1$ are 0.27 and 0.41, respectively.
The infinite sum, which appears in both equations~\ref{gencolabun} and \ref{gendeflect}, is a geometric series 
and converges to
\begin{equation}
 \sum_{l = 0}^{\infty} \left( \frac{r_0\nu_0}{H} \right)^l = \left( 1 - \frac{r_0\nu_0}{H} \right)^{-1}
\end{equation}
but only when 
\begin{equation}\label{crudecond}
 \frac{r_0\nu_0}{H} < 1 .
\end{equation}
It is interesting to note that the condition for the convergence of this series is very close to the condition 
for a ray to traverse the atmosphere, shown in equation~\ref{condition}. Indeed, they only differ by a factor 
$(1 + \nu_0)$ when evaluated at $r_0$. Since $(1 + \nu_0) \gtrsim 1$, equation~\ref{condition} is automatically satisfied
when equation~\ref{crudecond} is satisfied. The fact that they are not exactly the same is probably due to 
our crude method of estimating the remaining terms.

\section{Numerical model}\label{numerical}

\subsection{Ray-tracing algorithm}

We have made substantial improvements to an algorithm that, given a 1-D model atmospheric
temperature-pressure altitude profile, traces rays through that atmosphere and 
computes the number density column abundance along the rays as well as their deflection 
by the atmosphere. This algorithm, which has been previously described by 
B\'etr\'emieux \& Kaltenegger (2013, 2014) and references therein, suffered from 
numerical instabilities in the computed ray deflection, which started at densities of 
about $5\times10^{-5}$~amagat, and became worse with lower densities, such that the algorithm 
was useless for densities below $3\times10^{-6}$~amagat.
These numerical instabilities were not only due to improper treatment of integer to double precision
conversion of variables, but also intrinsically due to the method. Indeed, we were computing the ray 
deflection by first computing the azimuthal travel of the ray through the atmosphere, $\Delta\phi$ 
(see Fig.~\ref{fig1}), and then using 
\begin{equation}
 \omega = \Delta\phi + 2\theta_0 - \pi ,
\end{equation}
thus subtracting two large numbers from each other to get a much smaller quantity. 
We have modified the algorithm and reduced these numerical instabilities considerably. 
In this section, we will describe the new algorithm which we name MAKEXOSHELL.

The inputs to MAKEXOSHELL include a 1-D model atmosphere's temperature and pressure, or $T$ and $P$, as a 
function of altitude, $z$. Also required are the radius of the planet, $R_P$, the highest and lowest altitudes 
of the computational region, the number of rays and layers sampling this altitude region, and the 
STP refractivity of the atmosphere. Given temperatures and pressures from the model atmosphere, typically in 
increments of 1~km altitude, MAKEXOSHELL uses the ideal gas law to determine the corresponding densities. This 
serves as a basis for the rest of the computation done on a much finer altitude grid, or computational grid. 

MAKEXOSHELL assumes that the atmosphere within a model atmosphere layer follows an exponential behaviour 
with a constant density scale height as displayed in equation~\ref{expoatmos}. Once the 
scale height within that layer is computed, using the densities at its upper and lower boundaries, densities
are interpolated on the computational grid within that layer. Thus, MAKEXOSHELL builds-up a density profile 
on the computational grid from the model atmosphere densities. This is then combined with the 
input $\nu_{STP}$, as per equation~\ref{refrac}, to compute the altitude-dependent refractivity
of the atmosphere in the middle of each computational layer.

Rays are spread uniformly in altitude, in terms of their grazing height, across the specified computational 
region. At each of these grazing heights, the density, $n_0$, and refractivity, $\nu_0$ are determined 
from which the impact parameter of each ray can be computed according to equation~\ref{raypath}. 
With the impact parameter of a ray, $b$, as well as the density and the refractivity of the atmosphere on 
the computational grid, it is possible to compute the column abundance of gas intercepted by a ray, 
as well as the ray's deflection through the atmosphere.

\begin{figure}
\includegraphics[scale=0.45]{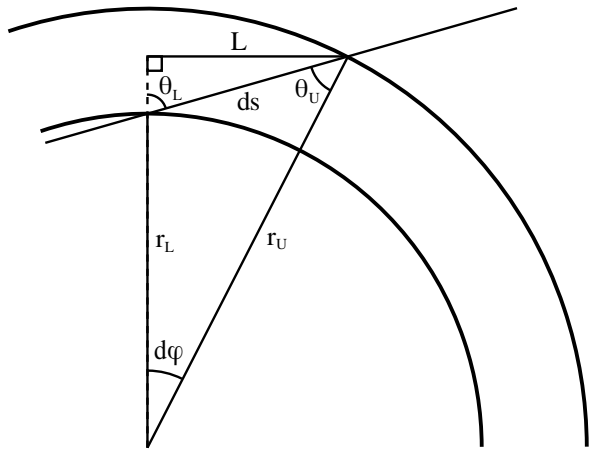}
\caption{Geometry of a ray travelling through an atmospheric layer with constant refractivity,
 i.e. in a straight line.
\label{fig2}}
\end{figure}

The computational grid is made up of layers that are thin enough that their densities, and hence 
their refractivities, are assumed constant across each layer. Thus, density and refractivity changes occur 
at their boundaries. In such a case, rays follow a straight line within each computational layer, 
as shown in Fig.~\ref{fig2}. We can see from this simple geometry that 
\begin{equation}
 d\phi = \theta_L - \theta_U ,
\end{equation}
and
\begin{equation}\label{geom}
L = r_U \sin d\phi = ds \sin\theta_L ,
\end{equation}
where $d\phi$ and $ds$ are the azimuthal travel of the ray and the path length of the ray within the 
computational layer, respectively. Subscript $U$ and $L$ denote quantities evaluated at the upper and 
lower boundary of the computational layer, respectively. Quantities without a subscript are evaluated 
in the middle of the computational layer, except for the impact parameter and the scale 
height. The former is ray-dependent, while the latter is determined by the model atmosphere 
layer to which the computational layer belongs. Both $ds$, given by
\begin{equation}\label{pathlength}
ds = \frac{r_U \sin ( d\phi )}{\sin\theta_L} = \frac{r_U \sin(\theta_L - \theta_U)}{\sin \theta_L} ,  
\end{equation}
derived from equation~\ref{geom}, and $d\phi$ depend on the zenith angles of the upper, $\theta_U$, 
and lower boundaries, $\theta_L$, of the computational layer. These are computed with
\begin{equation}
 \sin\theta_U = \frac{b}{(1+\nu)r_U} ,
\end{equation}
and
\begin{equation}
 \sin\theta_L = \frac{b}{(1+\nu)r_L} , 
\end{equation}
which have been derived from equation~\ref{raypath}.

The column abundance intercepted by a ray in a computational layer is derived 
simply by
\begin{equation}
 dN = n ds 
\end{equation}
while its deflection is given by 
\begin{equation}\label{incromega}
d\omega = \frac{\nu}{1+ \nu} \left( \frac{r}{H} \right) d\phi .
\end{equation}
Equation~\ref{incromega} is derived from equation~\ref{incdef}, recognising that 
\begin{equation}
 d\phi = \frac{\tan\theta}{r} dr ,
\end{equation}
which is the geometric term in equation~\ref{gradzenith}.
MAKEXOSHELL integrates $ds$ and $d\phi$ across all computational layers traversed by each rays, 
in order to get the column abundance and the deflection across an upward or downward pass for each
rays. The total column abundance and deflection experienced by a ray is twice that quantity
since the rays goes through these layers twice: once moving downward, and once moving 
upward through the atmosphere. 

Although the computational grid used in B\'etr\'emieux \& Kaltenegger (2013, 2014) was composed of 
0.1~km-thick layers, we have found that the precision of the computations improves significantly 
when the thickness of the computational layers decreases (see section~\ref{comparison}), and that 
the desired thickness depends on the atmospheric scale height. Hence, we now adjust this parameter 
as required by the atmosphere for which we do the computations. 

The contributions to the column abundance, $dN_{top}$, and ray deflection, $d\omega_{top}$, from 
atmospheric layers above the top of our computational altitude region are computed by assuming 
that these upper atmospheric regions are compressed into a shell one scale height thick. At these 
low densities, the ray trajectory is essentially straight, so that the path length of the ray, $ds_{top}$, 
inside this shell, is given by
\begin{equation}
 ds_{top} =  \sqrt{R^2_{top} \cos^2{\theta'_0} + 2H_{top}R_{top} + H^2_{top}} -R_{top} \cos{\theta'_0}
\end{equation}
where the subscript $top$ refers to quantities evaluated at the top of our computational region, or top 
of the atmosphere, with a corresponding refractivity, $\nu_{top}$, and zenith angle, $\theta_0'$
(see also Fig.~\ref{fig1}). Then, the contribution from these upper layers to the column abundance
is simply given by
\begin{equation}\label{coltop}
 dN_{top} = n_{top}ds_{top} ,
\end{equation}
and to the deflection of the ray by
\begin{equation}\label{defltop}
 d\omega_{top} = \nu_{top} \left( \frac{ds_{top}}{H_{top}} \right) \sin{\theta'_0} . 
\end{equation}
The latter equation was derived by using the small angle approximation in equation~\ref{pathlength}
to derive $d\phi$, which is then substituted into equation~\ref{incromega}.

\subsection{Refractivities}\label{refractivities}

One of the critical parameters for computing the deflection of the ray as well as the
column abundance along the refracted ray is the STP refractivity of the atmosphere, 
which depends on the STP refractivity of its major chemical species 
(see equation~\ref{refrac}). In this paper, we use the same refractivities for 
N$_2$, O$_2$, CO$_2$, and Ar, as in Table~2 from \citet{YB_LK_2013}. Note that this 
table has a small error: the STP refractivity for O$_2$ from \citet{bates1984} can be 
computed for wavelengths above 0.546~\micron, contrary to what this table states.
We are also considering two more species: H$_2$ and He. 
The STP refractivity of Helium is computed with the expression from \citet{M_P_1969}, 
described in \citet{Weber_2003}. For molecular hydrogen, we start from the formulation 
by Ford \& Brown (1973) of the H$_2$ Rayleigh cross section as a function of its
rotational quantum number J, from which we compute the cross section, $\sigma_R$,
for an equilibrium H$_2$ population at 300~K. We then determine its STP refractivity 
by using Rayleigh's formula, 
\begin{equation} \label{refractivity}
\nu_{STP} = \frac{n_{STP}}{w^{2}}\sqrt{\frac{3\sigma_{R}}{32\pi^{3}}} ,
\end{equation}
where $w$ is the wavenumber of the radiation. 

Since the refractivity of molecules hardly changes 
in the near-infrared (NIR) and infrared (IR) part of the spectrum, we can pick a 
single refractivity to represent that entire spectral region with minimal errors. In this paper, 
we use the refractivity of various molecules at 1~\micron~(shown in Table~\ref{table1}). 
Note that the derivation of the refractivity of H$_2$ from \citet{F_B_1973} includes a 
very slight temperature dependence (about a 0.3 per cent change in refractivity 
from 273.15 to 2000~K), but we will ignore it here and use the refractivity 
in Table~\ref{table1}.

We can compute the STP atmosphere refractivity for an
Earth-like atmosphere, considering the contribution of N$_2$, O$_2$, Ar, and CO$_2$, using the atmosphere 
composition listed in \citet{L_F_1998}. For a Jovian-like atmosphere, we use the mole fraction of helium
(0.1357) measured by the Galileo probe, also listed in \citet{L_F_1998}, and assume that the remainder of 
the atmosphere is composed of molecular hydrogen.

\begin{table}
 \centering
  \caption{STP refractivities of various gases and planetary atmospheres at 1~\micron~(see section~\ref{refractivities} for 
references).}
  \begin{tabular}{@{}cc@{}}
  \hline
   Gas/Atmosphere    &    $\nu_{STP}$ \\
 \hline
 H$_2$ & $1.37\times10^{-4}$ \\
 He & $3.48\times10^{-5}$ \\
\hline
 N$_2$ & $2.95\times10^{-4}$ \\
 O$_2$ & $2.68\times10^{-4}$ \\
 Ar & $2.79\times10^{-4}$ \\
 CO$_2$ & $4.43\times10^{-4}$ \\
\hline
 Jupiter & $1.23\times10^{-4}$ \\
 Earth & $2.90\times10^{-4}$ \\
\hline
\end{tabular}\label{table1}
\end{table}

\subsection{Atmospheric models}\label{atmmodels}

MAKEXOSHELL requires as input an atmospheric temperature and pressure profile as a function of altitude. 
In this paper, we only consider homogeneous isothermal atmospheres, so that both the temperature, $T$, and the 
composition are constant with altitude, which is currently done by some exoplanet atmosphere retrieval algorithm 
(see e.g. Charbonneau et al. 2009; Howe \& Burrows 2012; Anglada-Escud\'e et al. 2013; Ehrenreich et al. 2014; 
Waldmann et al. 2014).

\begin{table*}
 \centering
 \begin{minipage}{150mm}
  \caption{Input and derived model atmosphere parameters for various planetary systems}
  \begin{tabular}{@{}lcccccccc@{}}
  \hline
                   & \multicolumn{2}{c}{Earth} & \multicolumn{2}{c}{temperate Jupiter} & \multicolumn{2}{c}{HD189733b} & \multicolumn{2}{c}{GJ1214b} \\
    Parameters     & A & B &  C & D & E & F & G & H \\
 \hline
 Input &    &    &    &    &   &  \\
 \hline
 Composition \footnote{E: Earth-like; J: Jovian-like; See section~\ref{refractivities}} & E & E & J & J & J & J & J &  J \\
 T (K) & 255 & 255 & 255 & 255 & 1090 & 1090 & 550 & 550 \\
 $\Delta Z_{atm}$ (km) & 200 & 200 & 1000 & 1000 & 4500 & 4500 & 8800 & 8800 \\
 $\Delta$z (km) & 0.01 & 0.01 & 0.01 & 0.01 & 0.05 & 0.05 & 0.05 & 0.05 \\ 
 $\Delta$z$_{samp}$ (km) & 1 & 1 & 5 & 5 & 20 & 20 & 30 & 30 \\
 $\Delta$z$_{comp}$ (km) & 0.01 & 0.01 & 0.05 & 0.05 & 0.05 & 0.05 & 0.1 & 0.1 \\ 
 Gravity \footnote{V: varies with altitude; C: constant; See section~\ref{atmmodels}} & C & V & C & V & C & V & C & V \\
\hline
 Derived &              &      &      &      &         & & &  \\
\hline
 H (km) & 7.4 & 7.4 - 7.9 & 35.7 & 35.7 - 36.7 & 174.5 & 174.5 - 194.6 & 239.7 & 239.6 - 544.0 \\
 $z_{s}$ (km) & -17.07 & -17.12 & -82.22 & -82.32 & -401.8 & -403.8 & -551.85 & -568.90 \\
\hline
\end{tabular}\label{newtable}
\end{minipage}
\end{table*}

To build the model atmosphere, we start by specifying the mass of the planet, $M_P$, the surface pressure, $P_s$, 
the reference pressure, $P_R$, at the planetary radius, $R_P$, the thickness of the atmosphere, $\Delta Z_{atm}$,
and the isothermal temperature, $T$, of the atmosphere. 
We compute the mean molecular mass, $m$, from our specified bulk atmospheric composition.
We discretize our atmosphere in small altitude increments, $\Delta z$. We 
then determine the pressure scale height, $H'_i$, at each altitude by
\begin{equation}
 H'_i = \frac{kT}{mg_i}
\end{equation}
where the gravitational acceleration, $g_i$, can either vary with altitude
\begin{equation}
 g_i = g_s \left( \frac{R_P + z_s}{R_P + z_i} \right)^2 ,
\end{equation}
or remain constant at the surface gravity, $g_s$. The latter is computed by
\begin{equation}
 g_s = \frac{GM_P}{(R_P + z_s)^2}
\end{equation}
where $G$ is the gravitational constant, and $z_s$ is the altitude of the surface.

We then compute the pressure as a function of altitude using 
\begin{equation}
 P_{i+1} = P_i e^{-\Delta z/H'_i} , 
\end{equation}
starting with the surface at $i = 0$, and using the result of each iteration as input to the next, to the 
top of the atmosphere. The model atmosphere that is passed on to MAKEXOSHELL is sampled every $\Delta z_{samp}$, 
a much larger value than the thickness of a layer in MAKEXOSHELL's computational grid, $\Delta z_{comp}$.
Our choices of the vertical resolution of the computational grid and of the model sampling grid 
are adapted to the scale height of the planetary atmosphere. Typically, on the order of 1000 computational layers 
fit inside one scale height, and the resulting model atmosphere is sampled roughly every 100 layers. 

We have chosen not to include the centrifugal force for several reasons. The most important one is that
it requires knowledge of a planet's rotation period, which for exoplanets is usually unknown.
Coupled with the fact that the orientation of the spin axis with respect to the observer is 
also unknown, one end up with several free parameters that can not be unambiguously determined. 
Since the effect of the centrifugal force is to reduce gravity preferentially at a planet's equator, 
and hence increase an atmosphere's scale height, that can not be told apart from a local increase in 
temperature or a decrease in the mean molecular mass of the atmosphere. Hence, considering 
the effects of the centrifugal force adds a level of complexity that is unwarranted given 
the aim of this paper.

The isothermal temperature is chosen to be the mean planetary emission temperature, $T_e$. 
This is given by 
\begin{equation}\label{equte}
 T_e = \left[\frac{1}{4} (1 - \Lambda_B) \right]^{1/4} \left(\frac{R_*}{a} \right)^{1/2} T_*
\end{equation}
where the yet undefined parameters are the semi-major axis of the planetary orbit, $a$, the Bond 
albedo of the planet, $\Lambda_B$, and the effective temperature of the host star, $T_*$. This emission temperature 
is appropriate for a planet which thermally re-emits from its entire surface the absorbed stellar energy, 
and is often computed for various exoplanets over a breadth of possible Bond albedo, from 0 to 0.75
(see e.g. Charbonneau et al. 2009 for GJ1214b). In our simulations, we are choosing $T_e$ for a Bond 
albedo of 0.30, which is similar to the Bond albedo of Earth and the Jovian planets in our Solar system.

Table~\ref{newtable} summarises the various input parameters of the model atmospheres considered in this paper.
For each planet, we have one model where the gravity varies with altitude and one where it is constant, with 
otherwise identical input parameters. All of the model atmospheres have a surface pressure of 10~atm (atmosphere), 
and a reference pressure of 1~atm. The atmospheric thicknesses have been chosen such that the top of the 
atmosphere where gravity varies with altitude have densities ranging between $10^{-12}$ and $10^{-10}$~amagat.
The table also lists the corresponding altitude of the surface, and the range of density scale heights through
the atmosphere.

\section{Results and discussions}\label{validation}

\subsection{Validation of analytical expressions and ray tracing algorithm}\label{comparison}

The column abundance from our analytical expressions, $N_a$, is
\begin{equation}
 N_a = N_0 + N_1 + N_2
\end{equation}
expressed in equations~\ref{col0}, \ref{col1} and \ref{col2}, and the analytical ray deflection, $\omega_a$,
is 
\begin{equation}
 \omega_a = \omega_0 + \omega_1 + \omega_2
\end{equation}
expressed in equations~\ref{defl0}, \ref{defl1}, and \ref{defl2}. The outputs from MAKEXOSHELL 
can include, or not, the contribution from the upper layers (See equations~\ref{coltop} and \ref{defltop}). 
Comparison of one quantity, $Q_1$, relative to another, $Q_2$, is expressed by the percentage difference, 
$100(Q_1-Q_2)/Q_2$, either in terms of the absolute value or not.

To verify the accuracy of our ray tracing algorithm, we compare the column abundance, $N_n$, and ray 
deflection, $\omega_n$, obtained with MAKEXOSHELL with those computed from our analytical expressions, 
for a planet with Earth's radius and bulk mass ($R_P=6371.01$~km; 
$M_P=5.9736 \times 10^{24}$~kg). We use atmosphere model~A, which obeys the same condition as the 
assumptions used to derive our analytical expressions: isothermal temperature profile, constant 
composition with altitude, and constant gravity with altitude.
The lower boundary (see section~\ref{lowerbound}) occurs at a density of 4.03~amagat (or pressure of 3.76~atm), 
at an altitude of $-9.82$~km.

\begin{figure}
\includegraphics[scale=0.55]{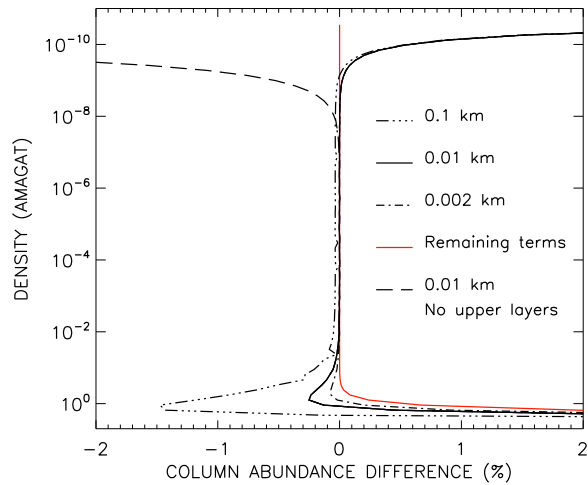}
\caption{Percentage difference of the numerical, $N_n$, relative to the analytical, $N_a$,
column abundance along a ray, as a function of the largest atmospheric density reached by that ray, 
for different thicknesses (0.1, 0.01, and 0.002~km - see inset legend) of MAKEXOSHELL's 
computational layers. That difference is also shown for a computational layer thickness of 0.01~km 
when the numerical contribution from the upper layers (see equation~\ref{coltop}) is not 
included (long-dashed line). The red curve shows our estimated 
contribution from the higher-order terms, $N_{rest}$ (see equation~\ref{gencolabun}).
\label{fig3}}
\end{figure}

\begin{figure}
\includegraphics[scale=0.55]{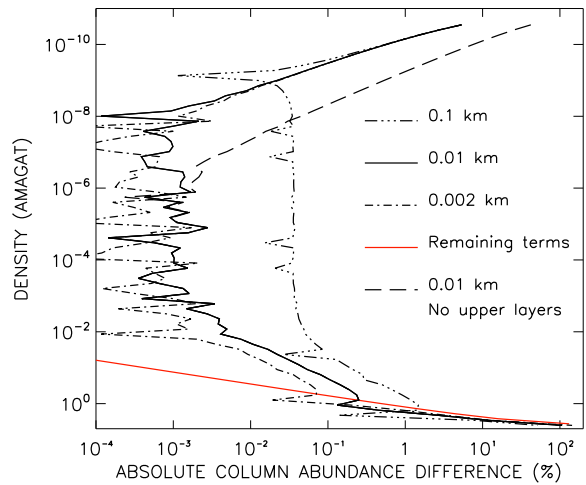}
\caption{Absolute percentage difference of the numerical, $N_n$, relative to the analytical, $N_a$,
column abundance along a ray, as a function of the largest atmospheric density reached by that ray, 
for different thicknesses (0.1, 0.01, and 0.002~km - see inset legend) of MAKEXOSHELL's 
computational layers. See the caption in Fig.~\ref{fig3}.
\label{fig4}}
\end{figure}

Figures~\ref{fig3} and \ref{fig4} show the difference of the numerical output of MAKEXOSHELL relative to 
our analytical derivation for the ray column abundance as a function of the largest atmospheric density 
reached by that ray. The nominal simulations use a thickness for the 
computational layers of 0.01~km and is shown by the solid line. For this vertical resolution, there is 
excellent agreement, better than 0.004 per cent, between our numerical algorithm and our analytical 
expressions over a wide range of densities (from about $4\times10^{-9}$ to $6\times10^{-3}$ amagat). 
At these densities, the numerical column abundances are slightly lower than those from the analytical 
expressions, and the obtained accuracy seems to be limited by numerical instabilities. 

At lower densities, significant errors occur near the atmospheric top boundary and the numerical 
algorithm finds larger column abundance than the analytical solution. This is due to our imperfect way of 
incorporating the contribution of the upper layers. However, not including it causes much larger errors 
(see the long-dashed line). One solution is to adjust the top boundary to confine these 
errors to atmospheric regions with densities low enough that they have negligible optical depths, and 
make no impact on the interpretation of remote sensing observations.

\begin{figure}
\includegraphics[scale=0.55]{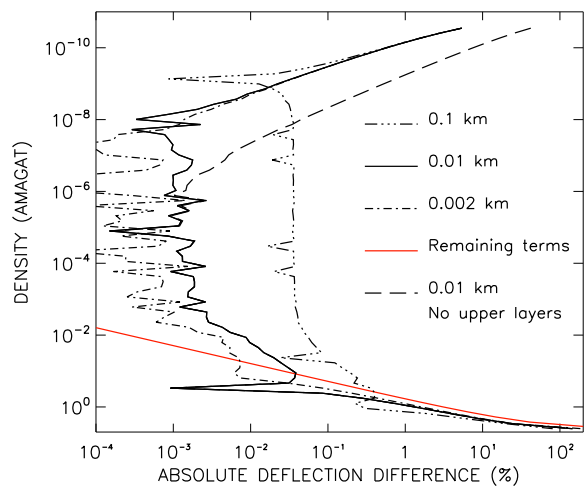}
\caption{Absolute percentage difference of the numerical, $\omega_n$, relative to the analytical, $\omega_a$,
ray deflection, as a function of the largest atmospheric density reached by that ray, 
for different thicknesses (0.1, 0.01, and 0.002~km - see inset legend) of MAKEXOSHELL's 
computational layers. That difference is also shown for a computational layer thickness of 0.01~km 
when the numerical contribution from the upper layers (see equation~\ref{defltop}) is not 
included (long-dashed line). The red curve shows our estimated contribution from the higher-order 
terms, $\omega_{rest}$ (see equation~\ref{gendeflect}).
\label{fig5}}
\end{figure}

At higher densities, the absolute difference in the column abundance increases and seems to follow a density 
power law, until it reaches about 0.3~per cent around 1~amagat. In these density regions, the numerical 
results are still lower than the analytical expressions. We have investigated 
whether this could be due to an inadequate computational grid resolution, and have done calculations for 
vertical computational layer thicknesses, $\Delta z_{comp}$, of 0.1 and 0.002~km (see the triple-dot dashed and the dot-dashed lines,
respectively, in Fig.~\ref{fig3} through \ref{fig5}). The computational grid resolution does indeed 
play a role, as the differences between numerical and analytical results decrease markedly when changing 
the layer thickness from 0.1 to 0.002~km. Our choice of 0.01~km is a compromise between accuracy and 
computational memory requirements, with 740 to 790 computational layers per atmospheric scale height.

For densities higher than about 1~amagat, the numerical results are higher than the analytical expressions
because the analytical expressions are missing the higher-order terms, and the difference increases 
dramatically with density as the lower boundary is approached. Although generally lower, it follows the same
trend as our crude order of magnitude estimate of the contribution from these higher-order terms (see 
equation~\ref{gencolabun}), shown by the red curve.

Differences between the numerical results and the analytical expressions of the ray deflection 
(see Fig.~\ref{fig5}) are generally higher than for column abundances, but follow roughly the same 
behaviour. The higher-order terms becomes non-negligible at densities starting around 
0.15~amagat and the difference in ray deflection is about 1.6 per cent at 1~amagat.

The excellent agreement between MAKEXOSHELL and our analytical expressions, in density 
regions where the higher-order terms are negligible, validates both methods simultaneously. 
However, the analytical expressions can only be used to compute the column abundance along the ray, 
and its deflection with good accuracy up to densities of about 1 and 0.15~amagat, respectively. 
On the other hand, MAKEXOSHELL can be used in denser regions where it captures qualitatively the 
expected location of a singularity near the lower boundary, as well as under more general conditions 
such as when gravity changes with altitude and the temperature profile is not isothermal. Indeed, 
MAKEXOSHELL's only assumption is that the atmospheric density scale height is constant within each 
layer of the input model atmosphere, which is approximately true if the vertical sampling of the 
model atmosphere is high enough.

Our analytical expressions are nevertheless very useful. They first provide a benchmark against 
which numerical models can be compared. Indeed, we converged on the current numerical algorithm and 
vertical computational resolution precisely because we could compare to our improved analytical 
expressions. They also allow derivation of analytical expressions for other quantities which depend 
on the column abundance or the ray deflection, as well as allow propagation of input 
parameter uncertainties, both of which are crucial for atmospheric retrieval algorithms.

\subsection{Relevance to temperate exoplanets}\label{relevance}

\subsubsection{Column abundance and ray deflection}\label{colanddef}

With the validation of our analytical expression and our numerical model, we can ask
to what degree does our analytical expressions and numerical model differ from the simplest 
analytical expressions that are used in some exoplanet atmosphere retrieval algorithm? 
The simplest analytical expressions are $n_0 \sqrt{2\pi Hr_0}$ for the column abundance, and 
$\nu_0 \sqrt{2\pi r_0/H}$ for the ray deflection. These are the leading term in our analytical 
expressions (see equations~\ref{seriescol} and \ref{seriesdefl}). We consider this question both 
for the Earth-like planet that we have already described, as well as for a temperate 
Jovian-like planet. 

Our temperate Jovian-like planet combines atmospheric model~C with Jupiter's radius 
and bulk mass ($R_P~=~69911$~km; $M_P~=~1.8986~\times~10^{27}$~kg), and orbits a Sun-like star at 
Earth's orbital distance. Since its density scale height is about five times that of 
our Earth-like planet, we can use a computational layer thickness of 0.05~km for MAKEXOSHELL and get 
similar agreements between the numerical results and our analytical expressions as for the Earth-like 
planet, because we have a comparable number of computational layers per scale height.

\begin{figure}
\includegraphics[scale=0.55]{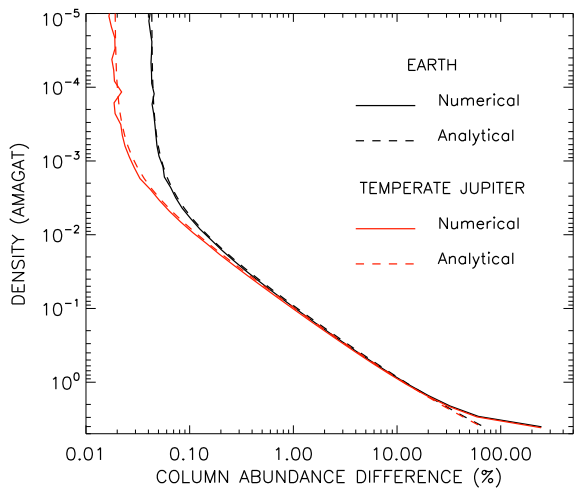}
\caption{Percentage difference of the numerical (solid line) and analytical (dashed line) ray 
column abundance relative to that from the analytical expression $n_0 \sqrt{2\pi Hr_0}$, 
as a function of the largest density reached by a ray, for Earth-like (black) 
and temperate Jovian-like (red, see text) planets.
\label{fig6}}
\end{figure}

\begin{figure}
\includegraphics[scale=0.55]{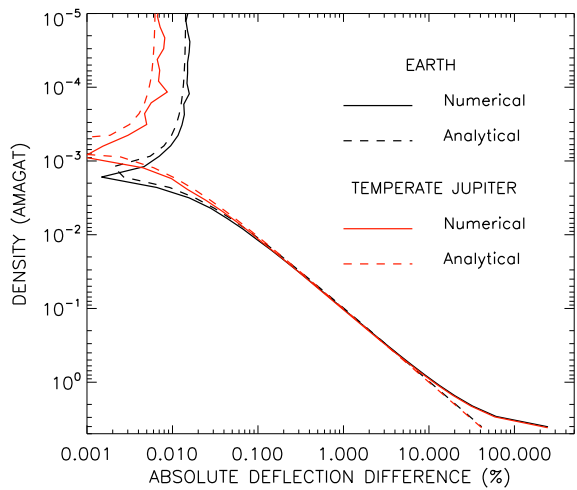}
\caption{Absolute percentage difference of the numerical (solid line) and analytical (dashed line) ray 
deflection relative to that from the analytical expression $\nu_0 \sqrt{2\pi r_0/H}$, as a function of the largest 
density reached by a ray, for Earth-like (black) and temperate Jovian-like (red, see text) planets. 
Note that the two minima around $10^{-3}$~amagat indicate a transition to negative differences which occur 
at lower densities. 
\label{fig7}}
\end{figure}

Figures~\ref{fig6} and \ref{fig7} show the percentage difference in the column abundance 
and the ray deflection, respectively, computed with our models, relative to those obtained 
with the simplest analytical expressions. This is shown both 
for our analytical expressions and our numerical algorithm, and for both type of planets 
considered. We are not showing the difference for densities lower than $10^{-5}$~amagat 
because it is constant, except close to the top boundary where these quantities are 
overestimated by MAKEXOSHELL (such as in figures~\ref{fig3} through \ref{fig5}).

The curves for both types of planets are remarkably similar except at densities lower 
than $10^{-2}$~amagat. The difference in those density regions comes from the 
$C_0$ and the $D_0$ coefficients in equations~\ref{seriescol} and \ref{seriesdefl}, which are 
power series of the ratio $(H/r)$. The similarities of these curves for both planets 
at deeper densities is fortuitous as the factor of 2 difference in $(r/H)$ is almost compensated 
by differences in refractivities, so that the dimensionless parameter $(r\nu/H)$
of these two planets is roughly the same at similar densities. Consequently, the lower boundary
occurs on the temperate Jupiter at a density of 4.15~amagat (or pressure of 3.87~atm), 
only marginally denser than on the Earth-like planet, at an altitude of $-52.32$~km.

For both planets, the simplest analytical expressions underestimate the column abundance and 
ray deflection by about 1~per~cent around densities of 
0.1~amagat ($\approx$0.1~atm), 10~per~cent around 1~amagat ($\approx$1~atm), and this error 
increases dramatically as rays approach the lower boundary, where $(r\nu/H)$
is equal to one.  Thus, in the denser atmospheric regions, the simple expressions are quite 
inaccurate. 

\subsubsection{Effective exoplanet radius}\label{effexorad}

Can these regions be observed in a transiting exoplanet, or are the optical depths there
so high that these differences have no impact on an exoplanet's effective radius? 
When stellar limb darkening is ignored, the effective radius, $R_{eff}$, of an 
exoplanet occulting the centre of its host star is given by 
\begin{equation}
 R_{eff}^2 = b_{max}^2 - 2 \int_{b_{min}}^{b_{max}} e^{-\tau} b db \label{effradius}
\end{equation}
where $\tau = \sigma N$ is the optical thickness traversed by a ray, $\sigma$ is an average atmospheric 
extinction cross section, and $b_{max}$ is the impact parameter of rays grazing the top of the atmosphere. 
$b_{min}$ is the impact parameter of rays grazing either the critical altitude or the planetary surface, 
whichever is larger. For a cloud-free homogeneous isothermal atmosphere, when one ignores refraction, 
the effective radius of an exoplanet occurs where grazing rays have traversed an optical thickness of 
about 0.56 (Lecavelier des Etangs et al. 2008). This result is sensitive roughly to four atmospheric scale 
height \citep{Griffith_2014}, where the atmospheric transmission $e^{-\tau}$ changes from 0.05 to 0.95, 
and column abundances are 5.4 and 0.1 times larger than at the effective radius, respectively.

As opacities change with wavelength, so does the effective radius of the exoplanet. A clear
Rayleigh scattering atmosphere free of other sources of opacity produces the smallest effective radius. 
The average Rayleigh cross section are about $4.0\times10^{-28}$~cm$^{2}$ and $7.5\times10^{-29}$~cm$^{2}$ 
at 1~\micron, for an Earth-like and Jovian-like composition, respectively. The smallest effective radius
at that wavelength then occurs where the column abundance along the ray are $1.4\times10^{27}$~cm$^{-2}$ 
and $7.4\times10^{27}$~cm$^{-2}$, respectively. The largest effective radius of a cloud-free Earth
in the NIR occurs in the core of a CO$_2$ band at $4.3~\micron$ at a column abundance of about 
$3.5\times10^{23}$~cm$^{-2}$ \citep{YB_LK_2014}, which we will use as a rough guideline of the largest 
value in the NIR of the effective radius of a planet for the discussion throughout this paper.
If one is interested in far ultraviolet spectral features, higher altitude regions of the atmosphere 
are actually of importance.

\begin{figure}
\includegraphics[scale=0.55]{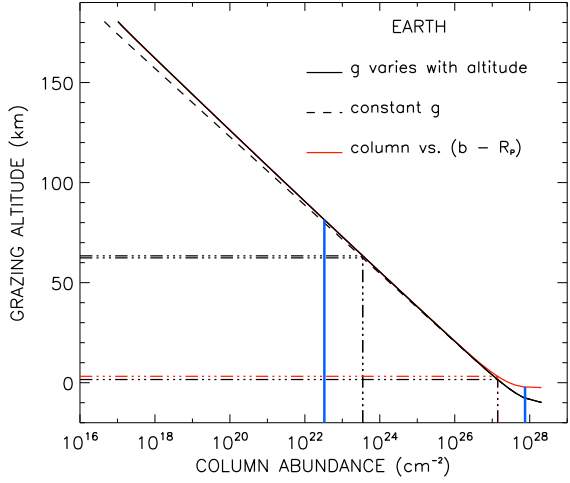}
\caption{Ray column abundance as a function of the grazing altitude of a ray, for the Earth-like 
isothermal atmosphere described in sections~\ref{comparison} and \ref{relevance}, when gravity 
varies with altitude (solid line), and when gravity is constant (dashed line). The red line is identical
to the solid line except that it is plotted as a function of the projected distance to the 
reference planetary radius with respect to the observer. The triple-dot-dashed lines bracket the 
likely effective radii of spectral features, and the vertical blue line bracket the range of column abundances 
to which these spectral features are sensitive (see section~\ref{effexorad}).
\label{fig8}}

\end{figure}

\begin{figure}
\includegraphics[scale=0.55]{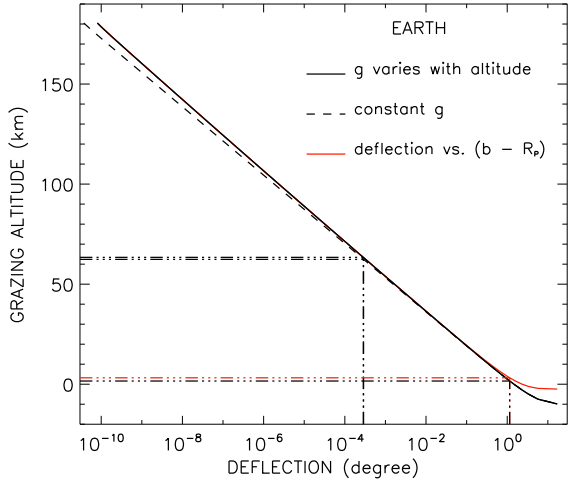}
\caption{Ray deflection as a function of the grazing altitude of a ray, for the Earth-like 
isothermal atmosphere described in sections~\ref{comparison} and \ref{relevance}. See the caption in Fig.~\ref{fig8}.
\label{fig9}}
\end{figure}

\begin{figure}
\includegraphics[scale=0.55]{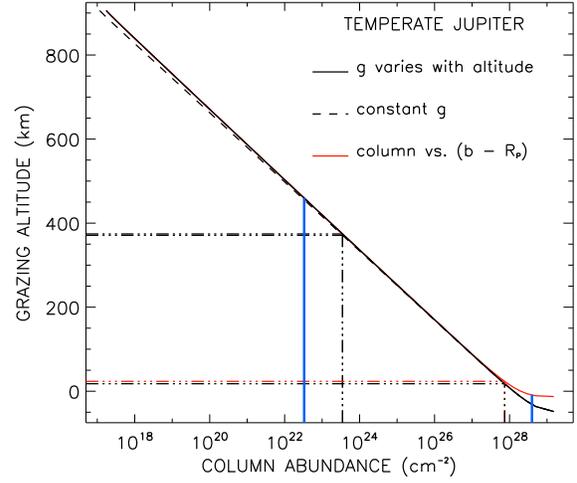}
\caption{Ray column abundance as a function of the grazing altitude of a ray, for the temperate Jovian-like 
isothermal atmosphere described in section~\ref{relevance}. See the caption in Fig.~\ref{fig8}.
\label{fig10}}
\end{figure}

\begin{figure}
\includegraphics[scale=0.55]{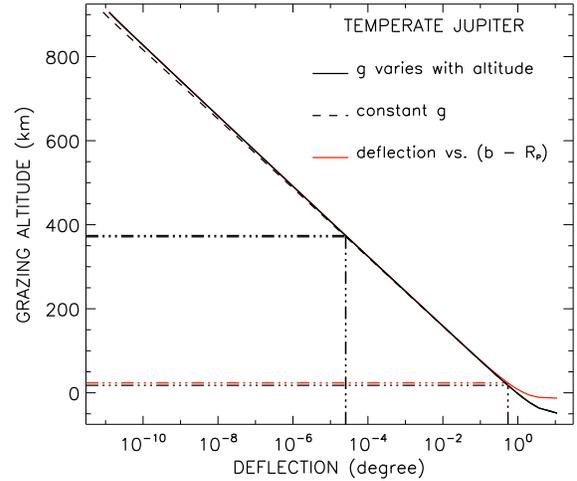}
\caption{Ray deflection as a function of the grazing altitude of a ray, for the temperate Jovian-like 
isothermal atmosphere described in section~\ref{relevance}. See the caption in Fig.~\ref{fig8}.
\label{fig11}}
\end{figure}

It is important to stress that the various column abundances quoted above are meant only as indicators to 
roughly translate the size of spectral features produced by Earth's atmosphere into equivalent features in 
other types of atmosphere, as well as get a feel for the range of atmospheric densities that contribute 
significantly to the creation of spectral features. They are in no way a substitute for actual radiative 
transfer calculations. In figures~\ref{fig8} through \ref{fig15}, the triple-dot-dashed lines bracket 
the grazing altitude corresponding to the smallest and largest effective radius using the simple column 
abundance criterion described above, as well as the range of ray deflection in that region.
The thick blue vertical lines bracket the range of column abundances to which these spectral features 
are sensitive, from about 10 times lower than at the largest effective radius to about 5.4 times larger 
than at the smallest effective radius. If our refractive model modifies substantially the column abundance
within this region, then the conditions for the validity of \citet{Lecavelier_2008}'s result breaks down
and the effective radius of the exoplanet will hence be modified. 

Figures~\ref{fig8} through \ref{fig11} show the ray column abundance and deflection computed with MAKEXOSHELL
for our isothermal Earth-like and temperate Jovian-like atmosphere, as a function of a ray's grazing altitude.
These are shown both for atmospheric models~B and D where gravity changes with altitude (solid line) and 
for models~A and C where it is kept constant at its surface value (dashed line). Note that in all these
simulations, the contribution from the upper layers are included.

For both of these temperate planets, the density scale height is much smaller than their radius, 
so that the variation of gravity with altitude does not make much of a change to the column abundances or ray deflection 
in the atmospheric regions that create spectral features. However, these regions include high density regions where 
our treatment of refraction computes column abundances which are at least 4 times larger than 
those computed with the simple analytical expressions used by \citet{Lecavelier_2008}. 
Hence, our treatment of refraction will modify the value of a transiting temperate exoplanet's effective radius 
at wavelengths where opacities are low.


\subsubsection{Effective versus actual scale height}\label{appscaleheight}

Furthermore, the effective radius depends on the variation of the column abundance along the ray with respect 
to its impact parameter, and not to the grazing altitude of a ray. When plotted versus the impact parameter 
(or in this case versus $b-R_P$, which is equivalent), both the column abundance and the ray deflection display a 
behaviour in the denser regions which is characteristic of a much smaller scale height (as shown by the solid red curve). 
The effective radius of the exoplanet reaches asymptotically a value, and thus effectively mimics a surface,
which is characteristic of a much smaller density than the highest density that can be probed. For our isothermal 
Earth atmosphere, the lower boundary, located at a density of 4.03~amagat, maps to an impact parameter 
2.4~km below our 1~atm reference. If one ignores this effect, one concludes that this impact parameter correspond 
to a density of only about 1.38~amagat. In the temperate Jovian-like atmosphere, the lower boundary, located 
at a density of 4.15~amagat maps to an impact parameter 12.7~km below our 1~atm reference, or an apparent density 
of 1.43~amagat.

These effects are due to the refractive deflection of the ray, and an altitude interval maps to 
increasingly smaller impact parameter interval for larger densities. Indeed this can be seen mathematically by differentiating 
equation~\ref{raypath} with respect to $r_0$, and then using equations~\ref{gradindex} evaluated at $r_0$ to obtain, 
\begin{equation}
 \frac{db}{dr_0} = 1 + \nu_0 \left( 1 - \frac{r_0}{H}\right) .
\end{equation}
A ray that traverses the atmosphere must satisfy equation~\ref{condition}, which results in $(db/dr_0) > 0$.
In the upper regions of the atmosphere where refractivities tend to 0, this derivative tends to 1, and 
altitude intervals are almost identical to impact parameter intervals.   
Near the lower boundary, this ratio tends to 0, so that an altitude interval 
maps to a vanishingly small impact parameter interval, thus forming what amounts to a refractive boundary layer. 

For our isothermal Earth atmosphere, the effective scale height of the 
observed column abundance has an average value of 4.3~km between 1.1 and 1.5~amagat, 1.8~km between 2.1 and 2.9~amagat, 
and only 0.32~km between 2.9 and 4.0~amagat, in contrast to the actual scale height of the atmosphere of about 7.4~km. 
A similar rapid decrease in the effective scale height occurs for the temperate Jovian atmosphere. 
When expressed in term of the planet's effective radius, the size of spectral features in transmission 
spectroscopy scale to first order with the effective scale height. This implies that features that originate in 
those dense region, such as those from dimers or collision induced absorption, will appear that much smaller and 
will be harder to detect.

\begin{table}
 \begin{minipage}{85mm}
  \caption{Given and derived parameters for HD189733b}
 \centering
  \begin{tabular}{@{}lcccc@{}}
  \hline
        & \multicolumn{3}{c}{References
  \footnote{B2005: \citet{Bouchy_2005}; T2008: \citet{Torres_2010}; S2010: \citet{Southworth2010}}}\\
  Parameters  & B2005 & T2008 &  S2010 \\
 \hline
 Given &              &            &    \\
 \hline
 R$_*$ (R$_\odot$)& 0.760 & 0.756 & 0.752 \\
 M$_*$ (M$_\odot$)& 0.820 & 0.806 & 0.840 \\ 
 T$_*$ (K) & 5050 & 5040 & 5050 \\
 R$_P$ (R$_J$) & 1.26 & 1.138 & 1.151  \\
 M$_P$ (M$_J$) & 1.15 & 1.144 & 1.150 \\
 a (AU) & 0.0313 & 0.0310 & 0.03142\\
\hline
 Derived &     &     & \\
\hline
 $\rho$ ($\times 10^3$ kg m$^{-3}$) & 0.763 & 1.030 & 1.000 \\
 $g_s$ (m s$^{-2}$) & 18.76 & 22.90 & 22.50 \\
 T$_e$ (K), $\Lambda_B = 0 $ & 1200 & 1200 & 1191\\
 T$_e$ (K), $\Lambda_B = 0.3$ & 1098 & 1098 & 1090\\
 T$_e$ (K), $\Lambda_B = 0.5$ & 1009 & 1009 & 1002\\
 T$_e$ (K), $\Lambda_B = 0.75$ & 848.6 & 848.8 & 842.5\\
 $\omega_c$ (\textdegree) & 7.572 & 7.504 & 7.381 \\
\hline
\end{tabular}\label{table2}
\end{minipage}
\end{table}

How much of an impact this effect has on the transit lightcurve depends on the part of the transit. When the planet 
occults the centre of its star, one can not observe atmospheric regions denser than the critical density of the 
planet \citep{YB_LK_2014}. During the course of the transit, the symmetry is broken, and one part of the planet's 
limb can be probed to higher densities while the opposite limb can only be probed to lower densities.
If the critical density of the planet occurs at a density where the effective scale height is not significantly 
different from the actual scale height, then this apparent decrease of the scale height of the denser regions
may not be observed, except possibly at the edges of the transit.

For a given column abundance, the Earth-like planet bends radiation more 
than the temperate Jovian-like planet. For instance, at a column abundance of $3.5\times10^{23}$~cm$^{-2}$, 
the ray deflection are about $2.8\times10^{-4}$ and $2.5\times10^{-5}$ degree, respectively, roughly a factor 
of 11 difference. The critical density to which one can probe the atmosphere 
during an exoplanet transit is higher for the temperate Jovian-like planet. Around a G2 star, the critical 
densities are 0.22 and 0.36~amagat for the Earth-like 
and the temperate Jovian-like planets, respectively. Around an M9 star (R$_* = 0.08$~R$_\odot$; T$_* = 2300$~K), 
the planet must have a semi-major axis of 0.0127~AU, in order to receive the same flux as a planet 1~AU from a G2 star. 
Since the planetary radii are comparable to the stellar radius, the critical deflections of the two 
planets are markedly different (1.87\textdegree and 3.79\textdegree, respectively) and so are the critical 
densities (1.32 and 3.00~amagat, respectively), which are high enough that the effective scale height of the 
atmosphere is significantly different from the actual scale height.

As our treatment of refraction makes a difference for temperate planets, we will now investigate 
whether this is also the case for hotter exoplanets which are more easily observed. As an example, we will consider
two planets that have already been repeatedly observed: the hot Jupiter HD189733b and the super-Earth GJ1214b.

\subsection{HD189733b}\label{HD189733b}

The NASA Exoplanet Archive \citep{Akeson_2013} lists three references for the system parameters of HD189733b 
and its host stars: its discovery by \citet{Bouchy_2005}, and further observations and reanalysis by
\citet{Torres_2010}, and \citet{Southworth2010}.
Table~\ref{table2} lists some of those parameters, as well as the mean mass density, $\rho$, surface gravity, $g_s$,
the mean emission temperature for various Bond albedo, $T_e(\Lambda_B)$, of the planet, and the critical 
deflection, $\omega_c$, of the planet-star system (see B\'etr\'emieux \& Kaltenegger 2014 for its definition), 
that we derive from these parameters. Note that we list the temperature for albedos between 0 and 0.75, the same 
range considered by \citet{Charbonneau_2009} and \citet{AE_2013} for GJ1214b.
Since its discovery, the planetary and stellar parameters of the HD189733 system have evolved quite substantially, 
but nevertheless the mean mass density and surface gravity of the planet from the parameters of \citet{Torres_2010} 
and \citet{Southworth2010} are very similar. What is very uncertain remains the temperature of the atmosphere, as 
this depends on the composition of minor species and whether clouds and aerosols are present. Hence, it is 
not worth to get into a discussion about whose planetary and stellar parameters are the best, as they do not 
dominate the uncertainties of the atmospheric properties, and we will merely use the revised parameters 
of \citet{Southworth2010} to get a sense of what effects are of importance. 

\begin{figure}
\includegraphics[scale=0.55]{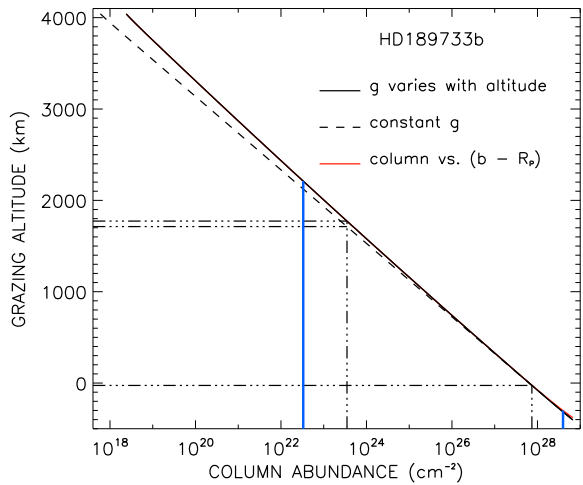}
\caption{Ray column abundance as a function of the grazing altitude of a ray for 
the Jovian-like isothermal atmosphere of HD189733b described in section~\ref{HD189733b}, 
using the S2010 parameters of Table~\ref{table2}. See the caption in Fig.~\ref{fig8}.
\label{fig12}}
\end{figure}

\begin{figure}
\includegraphics[scale=0.55]{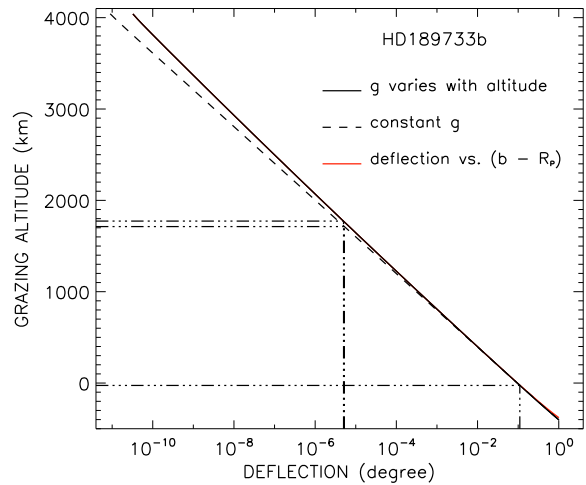}
\caption{Ray deflection as a function of the grazing altitude of a ray for 
the Jovian-like isothermal atmosphere of HD189733b described in section~\ref{HD189733b}, 
using the S2010 parameters of Table~\ref{table2}. See the caption in Fig.~\ref{fig8}.
\label{fig13}}
\end{figure}

We combine the planetary parameters (radius and mass) of HD189733b listed under S2010 in Table~\ref{table2}, 
with atmosphere models~E and F. Figures~\ref{fig12} and \ref{fig13}
show the ray column abundance and deflection as a function of the grazing altitude of a ray. 
Altitude regions where the effective scale height decreases dramatically is below the 
modelled regions. Just above our 10~atm ``surface'', the average effective scale height 
between 1.82 and 2.50~amagat is 146.1~km while the actual density scale height is 174.5~km.
The densities at a given pressure are about a factor of 4.3 lower than for our temperate planets because
the temperature is so much higher. Consequently, $(r\nu/H)$ never reaches unity, even at the ``surface''
where it is only about 0.14, and column abundance and ray deflection are increased by about 6 per cent 
by our treatment of refraction compared to the simplest analytical expressions.

The critical deflection of the planet-star system is about 7.4\textdegree (see bottom of Table~\ref{table2}), 
larger than the surface deflection of about 1.0\textdegree. Consequently, the critical 
altitude will occur below our ``surface'', and thus below the regions to which spectral features are 
sensitive (see blue vertical lines in Fig.~\ref{fig10}). Hence, refraction does not prevent 
any part of the atmosphere from being observed at wavelengths below 1~\micron. Although
Rayleigh scattering cross sections decrease with the fourth power of the wavelength, many molecular absorption
lines exist in the infrared, and are usually the dominant source of opacity.

Given our simple column abundance criterion (see section~\ref{effexorad}), the larger effective radius 
of the non-constant gravity model is 60~km larger than the constant gravity model, or about a third of a scale height, 
which translates into an extra stellar flux drop of $3.5\times10^{-3}$ per cent during transit. Since this is smaller 
than uncertainties in observations of HD189733b to date (see e. g. Fig.~4 in McCullough et al. 2014), this effect is 
not important for current exoplanet observations, but might be for future James Webb Space Telescope observations.

\subsection{GJ1214b}\label{GJ1214b}

The NASA exoplanet archives list a few references for the system parameters of GJ1214b and 
its host star: its discovery by \citet{Charbonneau_2009}, and further observations by 
\citet{Carter_2011}, \citet{Harpsoeetal_2013}, and \citet{AE_2013}. Table~\ref{table3} lists 
the given parameters from these different sources, as well as the mean mass density, surface gravity, 
average emission temperature, and critical deflection, that we derive from them (see section~\ref{HD189733b} 
for symbol definitions). Of these observations, \citet{AE_2013} uses recent parallax 
observations to re-evaluate the distance of the system, and combines it with new radial velocity 
measurements and transit observations to refine the system parameters. The derived planetary 
parameters are very similar to those previously obtained by \citet{Harpsoeetal_2013}, namely 
a low mean density and small surface gravity, but with higher temperatures than previously found. 
In this section, we will use their ``maximum probability'' planetary parameters (radius and mass) 
listed under AE2013 in Table~\ref{table3} with atmospheric models~G and H.

\begin{figure}
\includegraphics[scale=0.55]{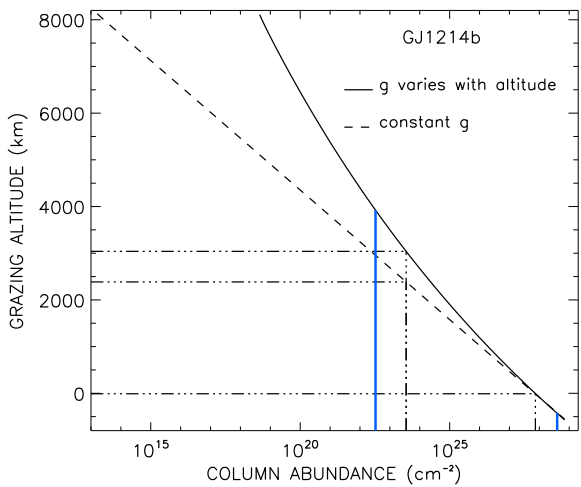}
\caption{Ray column abundance as a function of the grazing altitude of a ray for 
the Jovian-like isothermal atmosphere of GJ1214b described in section~\ref{GJ1214b}, using 
the AE2013 maximum probability parameters of Table~\ref{table3}. See the caption in Fig.~\ref{fig8}.
\label{fig14}}
\end{figure}

\begin{figure}
\includegraphics[scale=0.55]{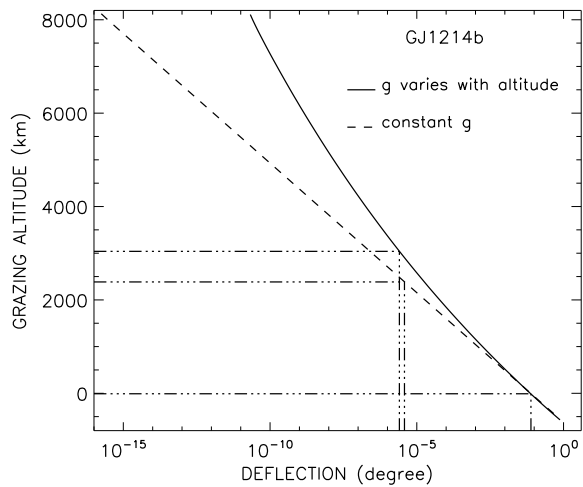}
\caption{Ray deflection as a function of the grazing altitude of a ray for 
the Jovian-like isothermal atmosphere of GJ1214b described in section~\ref{GJ1214b}, using 
the AE2013 maximum probability parameters of Table~\ref{table3}.
See the caption in Fig.~\ref{fig8}.
\label{fig15}}
\end{figure}

Figures~\ref{fig14} and \ref{fig15} show the ray column abundance and deflection as a function of 
the grazing altitude of a ray. At the ``surface'', $(r\nu/H)$ is 0.044, smaller than the value for HD189733b, 
and our treatment of refraction makes even less of a difference than for HD189733b. Indeed, column abundances 
at the ``surface'' are increased by about 3 per cent, and ray deflections by 1 per cent, by our treatment 
of refraction. The ray deflection at the surface is about 0.76\textdegree, far less than the critical deflection 
of 4.4\textdegree, and the critical altitude is below our ``surface''. Hence, for a Jovian composition, 
refraction does not prevent any part of the atmosphere of GJ1214b from being observed. 

\begin{table*}
 \centering
 \begin{minipage}{120mm}
  \caption{Given and derived parameters for GJ1214b}
  \begin{tabular}{@{}lcccccc@{}}
  \hline
        & \multicolumn{6}{c}{References
  \footnote{Ch2009: \citet{Charbonneau_2009}; Ca2011: \citet{Carter_2011}; H2013: \citet{Harpsoeetal_2013}
          ; AE2013: \citet{AE_2013}}}\\
    & Ch2009 & Ca2011 &  Ca2011 & H2013 & AE2013 & AE2013 \\
        &  & Method A & Method B & & Maximum & Expected \\
    Parameters    &  &  &  & & probability & value \\
 \hline
 Given &    &    &    &    &   &  \\
 \hline
 R$_*$ (R$_\odot$)& 0.211 & 0.210 & 0.179 & 0.216 & 0.213 & 0.211 \\
 M$_*$ (M$_\odot$)& 0.157 & 0.157 & 0.156 & 0.150 & 0.176 & 0.176 \\ 
 T$_*$ (K) & 3026 & 3170 & 3170 & 3026 & 3252 & 3252  \\
 R$_P$ (R$_E$) & 2.678 & 2.65 & 2.27 & 2.85 & 2.80 & 2.72 \\
 M$_P$ (M$_E$) & 6.55 & 6.45 & 6.43 & 6.26 & 6.26 & 6.19 \\
 a (AU) & 0.01439 & 0.01437 & 0.01225 & 0.01411 & 0.01449 & 0.01435 \\
\hline
 Derived &              &      &      &      &         & \\
\hline
 $\rho$ ($\times 10^3$ kg m$^{-3}$) & 1.88 & 1.91 & 3.03 & 1.49 & 1.57 & 1.70 \\
 $g_s$ (m s$^{-2}$) & 8.97 & 9.02 & 12.25 & 7.57 & 7.84 & 8.22 \\
 T$_e$ (K), $\Lambda_B = 0 $ & 559 & 584 & 584 & 571 & 601 & 601 \\
 T$_e$ (K), $\Lambda_B = 0.3$ & 511 & 535 & 535 & 522 & 550 & 550 \\
 T$_e$ (K), $\Lambda_B = 0.5$ & 470 & 491 & 491 & 480 & 506 & 506 \\
 T$_e$ (K), $\Lambda_B = 0.75$ & 395 & 413 & 413 & 404 & 425 & 425 \\
 $\omega_c$ (\textdegree) & 4.367 & 4.349 & 4.351 & 4.578 & 4.394 & 4.386 \\
\hline
\end{tabular}\label{table3}
\end{minipage}
\end{table*}

However, the combination of a high temperature with a low gravity and a 
low mean molecular mass creates a scale height which is a non-negligible fraction of the planetary 
radius. As a result, the variation of gravity with altitude makes quite a difference to the density 
altitude profile and to the resulting column abundances. Indeed, in order for atmospheric model~H
to have comparable densities at the upper boundary as models~B, D, and F, its atmospheric thickness must 
be significantly larger (see Table~\ref{newtable}). Ignoring the variation of gravity with altitude causes 
a drop in the column abundance of 5 orders of magnitude at that upper boundary, whereas it is less 
than 1~order of magnitude for the other models.

The size of spectral features increases by about 655~km when considering the effect of a non-constant
gravity with altitude, or $0.44$ per cent of the stellar radius, which is comparable to uncertainties in recent 
observations (see e.g. Fig.~4 in de Mooij et al. 2013, and references therein). However, one should then also 
expect spectral features to change the effective radius of the planet by about 3050~km, or about 2~per cent of 
the stellar radius, rather than the 0.6~per cent that has been observed to date.

We will not try here with our illustrative model to solve the mystery of GJ1214b's atmospheric composition 
but we will merely point out, as done by many previous analyses (see e.g. Bean et al. 2010; 
Croll et al. 2011; Crossfield, Barman \& Hansen (2011); Fraine et al. 2013; de Mooij et al. 2013; 
Kreidberg et al. 2014), that a haze or cloud 
layer is a possible solution. Indeed, the top of that layer would probably determine the lowest altitude 
that could be observed, rather than Rayleigh scattering, decreasing the altitude range available for formation
of spectral features. Incidently, our zero altitude reference pressure would also be correspondingly lower, 
shifting the densities in our atmospheric model to smaller heights.

\section{Summary and conclusions}\label{conclusion}

We have derived new analytical expressions for the column abundance along a ray
refracted by a planetary atmosphere with a constant density scale height, as well as 
the deflection experienced by that ray. These expressions were obtained by doing a 
Taylor power-series expansion of the exact integral formulation to the second order of 
the highest refractivity reached by the refracted ray. 
Surprisingly, the resulting analytical expression is not dependent on powers of the refractivity, 
but rather on powers of the dimensionless parameter $(r\nu/H)$. Although the refractivity is 
typically small, this dimensionless parameter is not necessarily so, and higher-order terms are
dominant in a temperate atmosphere when rays reaches densities larger than about 1 and 0.15~amagat
for the column abundance and the deflection, respectively. We have thus also derived an analytical 
expression that gives a rough estimate of the contribution from these higher-order terms. 

Since the scale height of most atmospheres are usually not constant, we have also developed a new numerical 
algorithm, MAKEXOSHELL, to trace rays through a spherically symmetric atmosphere given the STP refractivity of the 
atmosphere and an arbitrary one-dimensional temperature-pressure profile. We have compared the results of numerical 
simulations to values obtained with our analytical expressions for a 10~atm 255~K isothermal 
Earth-like atmosphere on a planet with a size and mass identical to Earth's. We have found that there is a 
better than 0.004 per cent agreement for densities up to $6\times10^{-3}$~amagat, both for the computed column 
abundance and the ray deflection, except near our upper atmospheric boundary where densities are extremely low, 
and in the denser regions where the higher-order terms dominate. The excellent agreement, between our 
analytical expressions and numerical results, validates both methods simultaneously.

We have built a few simple isothermal atmospheric models (temperate Earth-like planet, temperate Jovian-like 
planet, GJ1214b, and HD189733b) to determine the type of transiting exoplanets where our treatment of refraction, 
as well as the variation of gravity with altitude, makes an impact on their effective radius. The effects of a 
non-constant gravity with altitude are more pronounced for atmospheres where the atmospheric scale height is
not much smaller than the planetary radius. The combination of a hot temperature with a low gravity and mean molecular
mass, creates the proper condition in a Jovian-composition GJ1214b. If GJ1214b's atmosphere had a cloud or haze-free Jovian 
composition, the difference in density profile due to the variation in gravity
is sufficiently large that the resulting change in effective radius at high opacities is comparable to current 
observational uncertainties. It is far less important for HD189733b, and negligible for both temperate planets. 

The difference in the computed column abundance and the ray deflection, between our numerical treatment of refraction and
simple analytical expressions used in current exoplanet atmosphere retrieval algorithms, increases with the dimensionless 
parameter $(r\nu/H)$, and thus with the largest atmospheric density reached by a ray. We find that for hot Jupiters, such 
as HD189733b, as well as a Jovian-composition GJ1214b, atmospheric regions to which spectral features are sensitive are 
not dense enough for our improved treatment of refraction to make much of an impact. However, for temperate isothermal 
atmospheres of Earth-like and Jovian-like planets, the difference is about 1 per cent at 0.1~amagat, 10 per cent at 1~amagat,
and increase dramatically at larger densities, which can impact spectral features.

The column abundance and ray deflection tend toward infinite values, thus forming a refractive boundary layer,
as the deepest region that can be probed by transmission spectroscopy and stellar occultation is approached. This 
lower boundary occurs where the radius of curvature of a grazing ray exactly matches the radial position of 
the ray with respect to the centre of the planet, and is only dependent on the density profile of the planet's atmosphere. 
Rays that reach this boundary will spiral deeper into the atmosphere and be absorbed. The critical altitude 
discussed in \citet{YB_LK_2014}, can never be lower than this lower boundary no matter how close the exoplanet is 
to its star.

Combined with the bending property of the atmosphere, temperate atmospheric layers with densities larger than 1~amagat
have an apparent scale height significantly smaller than the actual density scale height of the atmosphere. This effective scale 
height tends to zero as rays approach the lower boundary. In transmission spectroscopy,
spectral features scale to first order with scale height when expressed in term of the effective planetary radius. 
Hence, spectral features that originate in these dense regions, such as those from dimers and collision-induced absorption, 
will be much harder to detect. Furthermore, this interesting effect effectively mimics a surface so that even temperate 
giant planets seem to have a surface in spectral regions where opacities are low. This effect can only be seen 
for exoplanets that are close enough to their star that the critical altitude of the planet-star system approaches 
this lower boundary. Given that it is easier to detect a planet by transit spectroscopy around cooler stars, and that 
as a result many exoplanet searches now specifically target M-dwarf stars, this is more likely to be indeed the case.

\section*{Acknowledgments}

The authors acknowledge support from DFG funding ENP Ka 3142/1-1.
This research has made use of the NASA Exoplanet Archive, which is operated 
by the California Institute of Technology, under contract with the National 
Aeronautics and Space Administration under the Exoplanet Exploration Program.


\label{lastpage}

\end{document}